# Creating Realistic Anterior Segment Optical Coherence Tomography Images using Generative Adversarial Networks

Jad F. Assaf, Anthony Abou Mrad, Dan Z. Reinstein, Guillermo Amescua, Cyril Zakka, Timothy Archer, Jeffrey Yammine, Elsa Lamah, Michèle Haykal, and Shady T. Awwad

*Abstract*— **This paper presents the development and validation of a Generative Adversarial Network (GAN) purposed to create high-resolution, realistic Anterior Segment Optical Coherence Tomography (AS-OCT) images. We trained the Style and WAvelet based GAN (SWAGAN) on 142,628 AS-OCT B-scans. Three experienced refractive surgeons performed a blinded assessment to evaluate the realism of the generated images; their results were not significantly better than chance in distinguishing between real and synthetic images, thus demonstrating a high degree of image realism. To gauge their suitability for machine learning tasks, a convolutional neural network (CNN) classifier was trained with a dataset containing both real and GAN-generated images. The CNN demonstrated an accuracy rate of 78% trained on real images alone, but this accuracy rose to 100% when training included the generated images. This underscores the utility of synthetic images for machine learning applications. We further improved the resolution of the generated images by up-sampling them twice (2x) using an Enhanced Super Resolution GAN (ESRGAN), which outperformed traditional up-sampling techniques. In conclusion, GANs can effectively generate high-definition, realistic AS-OCT images, proving highly beneficial for machine learning and image analysis tasks.**

## I. INTRODUCTION

Optical coherence tomography (OCT) is a non-invasive imaging technique used to capture high-resolution cross-sectional imaging in biological systems [1]. Anterior Segment OCT (AS-OCT), now routinely used in research and clinical practice, has proven itself as a major tool for anterior eye and cornea imaging [2].

Deep learning, a branch of machine learning, uses neural networks with multiple layers to discover intricate patterns within data. Generative Adversarial Networks (GANs) have recently emerged as a powerful tool for generating synthetic images that are indistinguishable from real images, termed "Deepfakes" [3]. GANs are a form of semi-supervised deep learning that includes two neural networks, a generator and a discriminator. The generator's purpose is to produce data that the discriminator can't distinguish from real data, and both networks learn and improve through competition [4]. GANs have many potential applications, including providing an alternative solution for expanding the quantity and variety of training data for data-hungry deep learning algorithms. [5, 6] These deepfakes can be altered to modify image characteristics such as decreasing OCT or fundus noise [7, 8], altering retina pigmentation to diversify and decrease bias in a dataset [9], and altering the disease severity of retinopathy of prematurity [10]. Many creative initiatives have emerged from GANs. One project improved fundus image resolution using a super-resolution GAN (SR-GAN) [11]. Another produced retinal angiography images from

This manuscript was submitted to IEEE-TMI on June 24, 2023 and received no sponsorship or financial support.

Jad F. Assaf, Anthony Abou Mrad, Jeffrey Yammine, and Elsa Lamah are with the Faculty of Medicine at the American University of Beirut, Lebanon (emails: jfa18@aub.edu.lb; ana69@mail.aub.edu; jey01@mail.aub.edu; eel01@mail.aub.edu).

Dan Z. Reinstein and Timothy Archer are with Reinstein Vision, London, UK, and London Vision Clinic, London, UK. Dan Z. Reinstein was also with Columbia University Medical Center, New York, NY, USA, Sorbonne Université, Paris, France, and Biomedical Science Research Institute, Ulster University, Coleraine, UK (emails: dzr@londonvisionclinic.com; timothy@londonvisionclinic.com).

Guillermo Amescua is with the Department of Ophthalmology, Bascom Palmer Eye Institute, University of Miami Miller School of Medicine, Miami, Florida (email: gamescua@med.miami.edu).

Cyril Zakka is with Department of Cardiothoracic Surgery, Stanford University, Stanford, USA (email: czakka@stanford.edu).

Michèle Haykal is with Faculty of Medicine, Saint Joseph University, Beirut, Lebanon (email: michelehaykal@gmail.com).

Shady T. Awwad (Corresponding Author) is with the Department of Ophthalmology, American University of Beirut Medical Center, Beirut, Lebanon (email: shady.awwad@aub.edu.lb).

Dr Reinstein is a consultant for Carl Zeiss Meditec (Carl Zeiss Meditec AG, Jena, Germany). Dr Reinstein is also a consultant for CSO Italia (Florence, Italy) and has a proprietary interest in the Artemis technology (ArcScan Inc, Golden, Colorado) through patents administered by the Cornell Center for Technology Enterprise and Commercialization (CCTEC), Ithaca, New York. The remaining authors have no proprietary or financial interest in the materials presented herein.

fundus photographs [12].

In the field of OCT imaging, previous studies have primarily focused on generating synthetic images of the retina [13, 14]. Top of Form These studies have explored the use of GANs for a variety of applications, such as post-treatment prediction of anti-VEGF treatment for Age-related Macular Degeneration [15, 16] and the removal of retina OCT shadows [17]. However, the number of studies that have targeted the anterior segment of the eye remains limited. One such study was aimed at generating AS-OCT B-scans for the detection of acute angle closure [18]. Another study generated Schiempflug tomography maps to augment a keratoconus dataset [19]. Despite these efforts, there is still a need for further research in this area to fully utilize the potential of GANs for anterior segment imaging.

In this paper, we propose a GAN-based approach to generate high-resolution and realistic AS-OCT images of the cornea. The GAN is trained on a dataset of real AS-OCT B-scans, and the synthetic images generated by the GAN are subjectively and objectively evaluated against real images in terms of visual quality and similarity. Additionally, we uncover image editing capabilities in the GAN's latent space, allowing us to discover meaningful image-editing vectors without the need for labeling. Moreover, we train an Enhanced Super Resolution GAN (ESGRAN) to upscale AS-OCT images without compromising their quality. The goal of this study is to demonstrate the potential of GANs in generating high-quality synthetic images of the cornea that can be used for various applications such as image editing or training and testing image analysis algorithms.

## II. METHODS

### A. Patient Selection and Data Preparation

This retrospective study was carried out at the American University of Beirut Medical Center (AUBMC) and received approval from the Institutional Review Board (IRB ID: BIO-2023-0021), in accordance with the principles of the Declaration of Helsinki.

Anonymized 16-mm and 10-mm wide AS-OCT B-scans from the MS39 machine (CSO, Florence, Italy) were retrieved from the AUBMC database for the period between September 2016 and July 2022. The 16-mm-wide scans on the MS-39 cover a depth of 8.5mm and can image the entire anterior segment, while 10-mm-wide scans cover a depth of 4mm and offer a detailed close-up view of the cornea. The images were then resized to 512 x 512 pixels using the bicubic interpolation on OpenCV.

To determine the distribution of our dataset, a refractive surgeon (S.T.A) assessed a random sample of 500 AS-OCT images from the AUBMC database and categorized them as normal scan, keratoconus, intracorneal ring segment, laser vision correction, corneal edema, intraocular lens, corneal transplant, corneal collagen crosslinking with visible demarcation line, Fuchs' endothelial dystrophy, cataract, implantable collamer lens (ICL), and Descemet's membrane endothelial keratoplasty.

### B. GAN Training

The GAN architecture chosen for this study was a Style-based Wavelet-driven Generative Model (SWAGAN) [20] which is based on the popular StyleGAN2 architecture [21] but operates on the frequency domain rather than the RGB (Red, green, and blue) domain of images for improved computing performance and visual fidelity. SWAGAN utilizes a style-based approach that disentangles the style and content information of the images, allowing for greater flexibility in generating images while maintaining the overall structure. Two separate GANs were trained for both image resolutions (16-mm and 10-mm) to ensure that each model was specifically tailored to its corresponding resolution, resulting in two distinct models. The GANs were trained for 500,000 iterations with a batch size of 4 and were optimized using Adam optimizer and a learning rate of 0.002. The generated images were normalized to the mean and standard deviation of the dataset. No augmentations were applied to the training set.

The GAN was implemented using PyTorch and training was parallelized over 2 RTX 3080 GPUs (NVIDIA, California, USA) donated by Hugging Face (New York, USA).

### C. GAN Selection

Model checkpoints and samples of generated images were saved for every 10,000 training iterations to monitor the training progression. The generator loss (path length regularization) and the discriminator loss (non-saturating loss with R1 regularization) were monitored during training for overfitting.

For every 10,000 iterations, the Fréchet Inception Distance (FID) score was calculated between 10,000 generated images and 10,000 randomly selected images from the real dataset. The model with the lowest FID score, indicating a higher level of similarity to real images, was selected as the best-performing GAN. FID is an objective comparison metric commonly used to evaluate the similarity of two sets of images, where a lower score (minimum of 0) indicates a higher level of similarity [22]. This metric not only assesses the visual similarity of the generated images to real images, but also takes into account their diversity and the similarity of their internal features, providing a more comprehensive and robust evaluation of the generated images as compared to relying solely on visual inspection.

Evaluation

To establish a benchmark for the Fréchet Inception Distance (FID) score achieved by the best-performing model, a comparison was made to the FID score between two randomly sampled sets of 10,000 images from the real dataset.

As a subjective evaluation of the GANs' performance, three refractive surgeons (S. T. A., D. Z. R., and G. A.) with over 15 years of clinical experience each were sequentially shown a total of 100 images, out of which 50 were randomly selected from real OCT scans, and 50 were randomly generated by the GANs. The surgeons were then asked to determine whether each image was real or fake. Two tests were created for both 16- and 10-mm GANs with 50 images each. To prevent bias, no information about the distribution of images was shared. All images were scaled to a uniform resolution (512 x 512 pixels) then were stretched back to their original aspect ratios of 512 x 937 pixels for the 16-mm images and 512 x 1320 pixels for the 10-mm images.

The results were recorded and analyzed, and the accuracy of the surgeons' evaluations was calculated as the percentage of correctly identified real and synthetic images. Additionally, sensitivity and specificity were calculated. Sensitivity indicates the proportion of real images correctly identified, while specificity indicates the proportion of fake images correctly identified. To determine if the accuracy of each ophthalmologist was significantly different from random guessing (50%), a binomial analysis was conducted using the SciPy library (v1.9.3) in Python (v3.10.6), with a significance level set at 0.05. Furthermore, the inter-reader agreement was evaluated using Fleiss' kappa test (StatsModels package v0.13.5) to compare the responses of the three ophthalmologists and assess the differences in their answers, and thus the validity of the GANs. In this method, the agreement between the raters is compared to what would be expected by chance alone. This method compares the agreement between the raters to what would be expected by chance alone. A kappa score of 0 indicates no agreement beyond chance, while a score of 1 indicates perfect agreement.

To demonstrate that the generated images are suitable for machine learning tasks, a simple convolutional neural network (CNN) model to classify images into ICL and non-ICL categories. In the first experiment, the CNN was trained on a dataset of 150 real ICL and 150 real non-ICL images. In the second experiment, the CNN was trained on a dataset of 150 ICL and 150 non-ICL images generated by our GAN model. In the third experiment, both datasets were pooled together to train the CNN. Model accuracy was recorded after 15 epochs on a test set of 100 real ICL and 100 real non-ICL images. The real OCT images used in this test were not part of the dataset used to train the GANs as to prevent data leakage.

### D. Leveraging GANs for AS-OCT Image Editing

Controlling and modifying the output of generative models is important in generating new data points. This is achieved through the use of latent space representation. In the context of GANs, a latent space is a representation of data and is a feature space of a generative model where the model is able to generate new, unseen data points [23]. The space is typically represented by a lower-dimensional vector, often referred to as a latent vector, and the generator is trained to generate new data points by sampling from this latent space.

GANs are frequently used as powerful image editors. We aim to uncover such capabilities by using Closed-Form Factorization, a technique that discovers meaningful directions in the GAN latent space [24]. Labeling OCT images to extract meaningful features is very time-consuming and impractical at a large scale. Therefore, by leveraging closed-form factorization, we can explore the latent space of the GAN in an unsupervised manner and identify directions that correspond to specific image editing capabilities, such as adding features of anterior segment diseases and ocular procedures (data augmentation) or to remove artifacts (image pre-processing). This approach allows us to discover image-editing vectors without the need for extensive manual labeling, making it scalable and efficient.

To identify meaningful directions in the latent space, we generated image samples and manipulated them in certain latent directions using Closed-Form Factorization. We then analyzed the resulting images to identify meaningful directions in the latent space. This process was done iteratively, by manipulating the latent variables and visually assessing multiple generated images, to ensure that meaningful directions in the latent space were identified that hold across multiple samples and generation seeds.

### E. Super Resolution

To bypass hardware limitations and further enhance the resolution of OCT images beyond the generated image resolution of 512 pixels, we aimed to upscale images using a super resolution GAN created with an Enhanced Super Resolution GAN (ESRGAN) [25]. ESRGAN is a specialized type of GAN that is designed to perform single-image super-resolution, which is the process of generating high-resolution images from low-resolution input images. ESRGANs have been widely used in various fields, including medical imaging, where they have shown promising results in improving the resolution and quality of medical images such as retinal fundus images [26] or optic nerve head OCT images [27].

To accomplish this, we created separate models for 10-mm and 16-mm scans using the same AS-OCT dataset. The ESRGANs were trained to upscale images from a downscaled resolution (2x) back to their original resolution. We first preprocessed the dataset by dividing each image into patches of size 128 x 128 pixels and filtering out those with a low mean signal to remove background and empty patches. We then fine-tuned pretrained ESRGANs for 500,000 iterations and selected the model with the most realistic results based on objective evaluation of a validation set using the Learned Perceptual Image Patch Similarity (LPIPS), which has proven correlation with human evaluation [28]. LPIPS measures the perceptual similarity between an image and its transformation, with lower LPIPS indicating higher similarity between two images. Finally, we compared our super resolution GANs with other traditional image upscaling techniques (bicubic interpolation, bilinear interpolation, nearest point interpolation, and Lanczos resampling) using the LPIPS metric.

## III. RESULTS

### A. Dataset

For this study, we collected a total of 85,196 16-mm AS-OCT B-scans from 2,905 eyes of 1,525 patients and 57,432 10-mm AS-OCT B-scans from 4,338 eyes of 2,360 patients. Inspection of 500 randomly selected OCT B-scans yielded the results summarized in table I. The AUBMC AS-OCT database included patients with a wide range of conditions and anterior segment procedures, including synthetic and allogenic intrastromal corneal rings segments (ICRSs), implantable collamer lenses (ICLs), and lenticule extractions. This diversity allowed the GAN to be trained on a varied dataset and generate similar realistic examples.

| AS-OCT CONDITION | PROPORTION (%) |
|---|---|
| NORMAL SCAN | 52.0 |
| KERATOCONUS | 19.2 |
| INTRACORNEAL RING SEGMENT | 11.2 |
| LASER VISION CORRECTION | 8.0 |
| CORNEAL EDEMA | 7.8 |
| INTRAOCULAR LENS | 5.2 |
| CORNEAL TRANSPLANT | 5.2 |
| FUCHS' ENDOTHELIAL DYSTROPHY | 4.8 |
| CORNEAL COLLAGEN CROSSLINKING WITH VISIBLE DEMARCATION LINE | 2.6 |
| CATARACT | 1.6 |
| IMPLANTABLE COLLAMER LENS | 1.2 |
| DESCEMET'S MEMBRANE ENDOTHELIAL KERATOPLASTY | 0.6 |

Table. I. Distribution of the OCT B-scans after inspection of 500 OCTs. Results do not add up to 100% as some conditions can be done concomitantly (e.g., ICL and ICRS).

### B. GAN Selection

The training process for both models was completed in 6 days each. Using the FID score, the models at iterations 160,000 and 40,000 were selected as the best for the 16-mm and 10-mm GANs, respectively. Figure 1 presents a sample of generated images from the 10- and 16-mm GANs, showcasing the variety of forms present in the images, including ICLs, corneal ring segments, corneal haze, and laser ablation, among others.

### C. GAN Evaluation

The FID score for the real images was found to be 1.10 (benchmark), while the scores for the generated images using the 16-mm and 10-mm GANs were 5.12 and 7.60 respectively. While there is no universal benchmark for the FID score, other work on GANs for OCT retina report FID values between 12.0 and 16.7 [29].

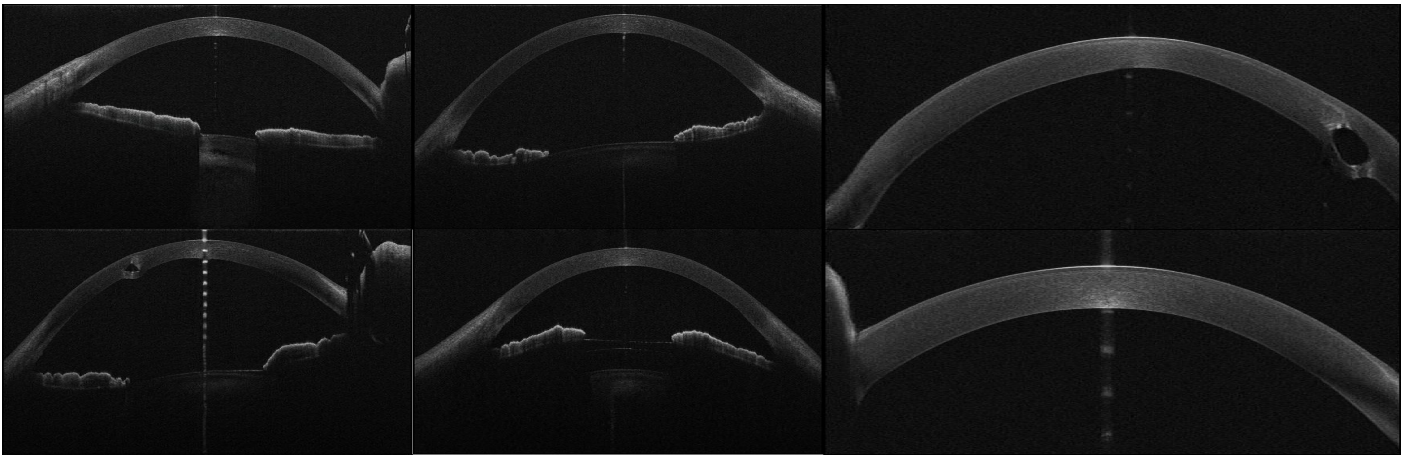

Fig. 1. Sample of generated images from the 10- and 16-mm GAN. We notice the variety of forms present. Some images have eyelid artifacts, others have corneal ring segments, or an ICL. The top right image has an irregular cornea (keratoconus) and has subepithelial haze while missing the bowman's bilaminar membrane due to laser ablation.

The three ophthalmologists achieved accuracies of 48%, 48% and 58% on the test for the 16-mm GAN and 48%, 52% and 56% on the 10-mm GAN. The full results with the sensitivity and specificity are illustrated in table II. For all evaluations, accuracy was statistically similar to random chance (50%) for all ophthalmologists (p-value>0.3). Fleiss' kappa scores of 0.196 and 0.047 were obtained for the 16- and 10-mm GANs, respectively, indicating that there was only a slight agreement among the three ophthalmologists beyond chance and that they are no better than random at detecting the generated scans.

|  | 16-MM | | | 10-MM | | |
|---|---|---|---|---|---|---|
|  | ACC | SEN | SPE | ACC | SEN | SPE |
| REVR 1 | 48% | 92% | 4% | 48% | 88% | 8% |
| REVR 2 | 48% | 64% | 32% | 52% | 96% | 8% |
| REVR 3 | 58% | 76% | 40% | 56% | 68% | 44% |

Table. II. Confusion matrix of expert assessment of synthetic versus real images. ACC: accuracy. SEN: sensitivity. SPE: Specificity. REVR: Reviewer.

The findings suggest that the generated images exhibit a high degree of similarity to the real images in terms of both visual appearance and internal features. This supports the conclusion that the GAN was able to generate high-quality synthetic images of the cornea. Additionally, we manually inspected 500 images from each GAN for obvious artifacts which are illustrated in Figure 2.

Furthermore, the CNN developed to classify normal from post-operative ICL OCTs was able to achieve similar performance using both real and synthetic data, with an accuracy of 78% over the real images, and 76% on synthetic images. With both datasets pooled together, however, the CNN achieved an accuracy of 100% on the test set, demonstrating that augmenting a dataset with GAN-generated images improves model training and performance. These findings suggest that the generated images can be used for machine learning tasks, at least in the context of ICL. It is worth noting that one of the advantages of GANs is in their ability to generate unlimited data when needed.

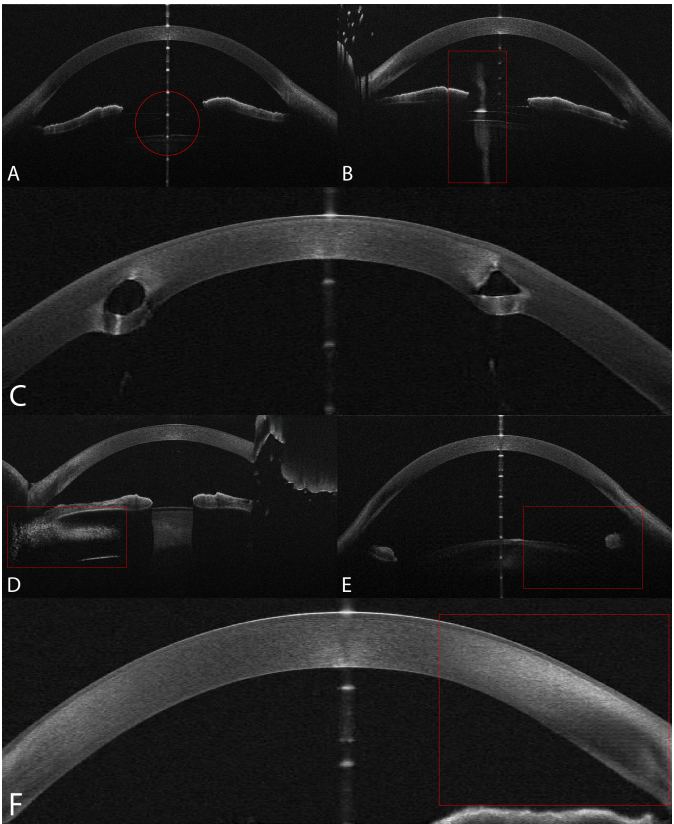

Fig. 2. Artifacts found in the generated AS-OCT images. A) Abnormal curvature noted in both the ICL and the cornea. B) Abnormal flaring of the ICL which is usually sharp and straight rather than diffuse. C) Two different ICRSs in 1 cornea which arguably does not happen in real life. D) Abnormal patch over the image that does not correspond to any structure. E) Grainy/patchy portion of the image that does not resemble the OCT speckle. F) Abnormal grainy linear cornea pattern.

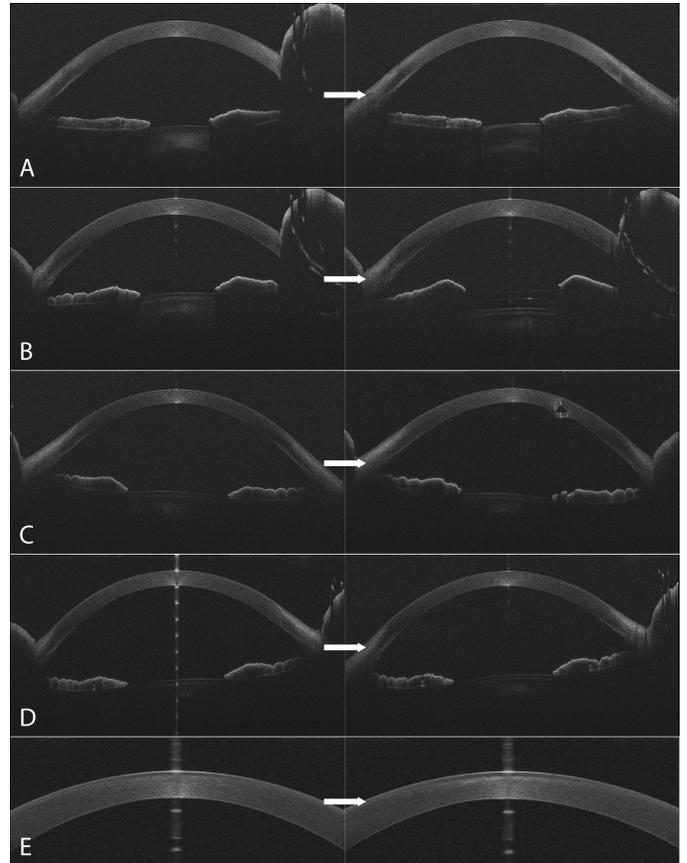

Fig. 3. Image editing using Closed-Form Factorization. Each row corresponds to an image transformation with leftmost figures as the original images, and the rightmost images as their corresponding edits. Images were transformed to A) remove eyelids, B) insert an ICL, C) insert a synthetic ring segment, D) remove the central flare artifact, E) and convert or simulate a PRK with myopic correction on a normal cornea, noticeable by a loss of the central bowman's bilaminar membrane and subepithelial haze.

### D. Image Transformation

We were able to identify directions in the latent space that correspond to specific AS-OCT image editing capabilities. We demonstrated this through the visualization of various image transformations in Figure 3. Some of the transformations include the ability to add or remove eyelids (Figure 3A), implantable collamer lenses (ICLs) (Figure 3B), and corneal ring segments (Figure 3C), eliminate central flare artifacts (Figure 3D), and even simulate photorefractive keratectomy (PRK) with myopic correction and subepithelial haze (Figure 3E). These results demonstrate the potential of closed-form factorization as a powerful tool for discovering the image editing capabilities of GANs. While vectors A, B, C, D, and E can serve as data augmentation techniques, vector D is useful for image pro-processing.

### E. Super Resolution

We trained a super resolution GAN using ESRGAN to upscale OCT images by a factor of 2. The selected 10- and 16-mm models scored a LPIPS of 0.0847 and 0.0962, respectively, on the validation images. These metrics compare favorably with different up-sampling techniques, the best of them, bicubic interpolation, scoring a higher and therefore worse LPIPS of 0.1167 and 0.7320 on the 10- and 16-mm images of the validation set, respectively.

Our approach effectively enhanced the resolution of OCT images with minor quality loss (Figure 4). Our method demonstrated superior results, characterized by sharper edges and more detailed structures, as compared to traditional interpolation techniques. Overall, our approach enabled us to upscale GAN-generated OCT images and enhance their resolution to reach 1024 x 1024 pixels.

## IV. DISCUSSION

In this study, we developed Generative Adversarial Networks (GANs) capable of generating realistic anterior segment optical coherence tomography (AS-OCT) scans of the cornea. We trained the GANs using a dataset of real AS-OCT B-scans, and subsequently evaluated the synthetic images generated by the GANs against real images in terms of visual fidelity and potential for machine learning use. To enhance the resolution of the generated images, we employed a super-resolution GAN for upscaling. The findings of this study revealed that the synthetic images generated by the GANs were visually similar to real images, to the extent that refractive surgeons had difficulty in distinguishing between them. Furthermore, we demonstrated the utility of the generated images for training machine learning algorithms. This work was inspired by StyleGAN2, an NVIDIA-produced GAN that generates realistic deepfakes of human faces [21].

One of the main advantages of GANs is their ability to generate synthetic images from unlabeled data, which allows them to capture the underlying structure within the data. In this study, we utilized a diverse unfiltered dataset of AS-OCT B-scans, which includes patients with a wide array of anterior segment conditions and procedures. This diversity allowed the generation of similar samples by the GANs. For instance, an interesting example illustrated in Figure 1 (top right OCT B-scan) corresponds to a postoperative keratoconic cornea that has undergone ICRS insertion with PRK. This is evident from the irregular posterior curvature and steepening characteristic of keratoconus, as well as the absence of the bowman's bilaminar membrane with subepithelial haze, indicative of PRK. Indeed, this is combination of procedures is common at AUBMC. Furthermore, we identified meaningful GAN image transformations in an unsupervised manner. Isolating such transformations can be useful to enhance a dataset with specific features (e.g., augment a laser vision correction dataset), or to preprocess an image (remove artifacts or decrease noise). In addition, we utilized a super-resolution GAN to effectively enhance the resolution of the generated AS-OCT images, surpassing the capabilities of traditional interpolation techniques.

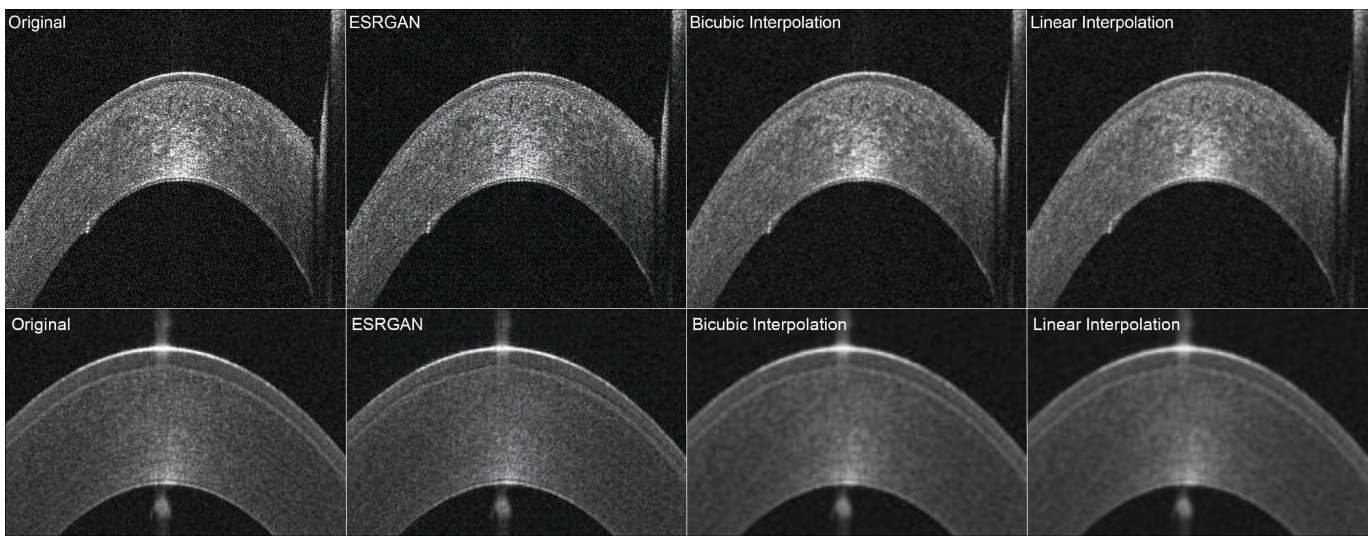

Fig. 4. Comparison of different upscaling techniques. AS-OCT images were zoomed in on the cornea to appreciate corneal morphology and detail. Original images are unaltered AS-OCT scans. The remaining images are the product of upscaling (2x) a twice downscaled version of the original image using the ESRGAN, the bicubic interpolation technique, and the linear interpolation technique. We notice sharper edges and detail in the ESRGAN compared to conventional techniques.

However, there are several limitations to our study that should be acknowledged. Firstly, the resolution of the generated images was limited to 512 x 512 pixels, which was then upscaled to 1024 x 1024 pixels using our super-resolution GAN. Although this resolution may be lower than what ophthalmologists typically work with (usually 1000 pixels and above), it is worth noting that for many machine learning tasks, high-resolution images are not necessary as they are often scaled down to reduce the computational requirements for processing [30]. Secondly, the continuous latent space learned by the GANs may not fully reflect the real-world attributes and pathophysiology of the cornea. While the synthetic images generated by the GANs may appear realistic, they may not accurately represent the underlying pathology of a real cornea. For example, Figure 3 illustrates a cornea generated by the GAN with a combination of two different ICRS types, which, while theoretically possible, arguably does not reflect a real-life occurrence. Similarly, transformations such as eyelid removal or ICRS insertion may appear realistic in the synthetic images, but they may not correctly capture the actual tissue characteristics underneath the eyelid or corneal shape and tomography changes after ICRS insertion.

Despite these minor drawbacks, GANs have clearly emerged as a highly valuable tool in the industry. Our experiments underscore the potential of synthetic data in augmenting datasets for artificial intelligence (AI) and deep learning, where large volumes of data are essential for effective model training. Notably, GANs have been substantiated to enhance training scores and accuracy, even in the domain of medical imaging [31], a field known for its data-intensive requirements. Furthermore, GANs offer significant advantages by generating synthetic data that circumvents legal, ethical, and technical complexities associated with medical records and real patient data. The use of medical data is often constrained by privacy concerns, data sharing agreements, and regulatory compliance requirements. GANs, by generating synthetic data that does not involve real patient data, offer a feasible solution to mitigate these challenges and provide researchers and practitioners with a valuable resource for model training without compromising patient privacy and data security.

In the future, specialized image-to-image GANs could potentially be applied to remove the central flare artifact and regenerate the distorted corneal tissue underneath. This is an area where GANs can be particularly useful, as the central flare artifact is a common problem that affects the quality of AS-OCT images where the angle of incidence of the light is zero. Although we have identified the potential for such a transformation vector (as shown in Figure 3D), the selected transformation vector was not entirely disentangled (pure) and altered some other minor image properties along the way, unrelated to the flare. Additionally, the transformed corneal properties may not accurately reflect the real corneal properties without the flare, and therefore, a separate study is needed to adequately address this issue. Another useful application is to reduce noise from the AS-OCT signal, therefore improving image quality. Similar techniques have been applied to retina OCT [32], and could potentially be adapted for AS-OCT images to decrease image capturing speed and enhance diagnostic capabilities.

## V. CONCLUSION

In conclusion, this study demonstrates the potential of GANs in generating high-quality synthetic images of the cornea, overcoming ethical concerns associated with real-world medical data. Despite the limitations of the study, the results are promising, and future research can build upon this work to further enhance the performance of GANs in specific application.

## REFERENCES

[1]  D. Huang *et al.*, "Optical coherence tomography," (in eng), *Science,* vol. 254, no. 5035, pp. 1178-81, Nov 22 1991, doi: 10.1126/science.1957169.
[2]  S.-H. Lim, "Clinical Applications of Anterior Segment Optical Coherence Tomography," *Journal of Ophthalmology,* vol. 2015, p. 605729, 2015/03/02 2015, doi: 10.1155/2015/605729.
[3]  J. S. Chen *et al.*, "Deepfakes in Ophthalmology: Applications and Realism of Synthetic Retinal Images from Generative Adversarial Networks," *Ophthalmol Sci,* vol. 1, no. 4, p. 100079, Dec 2021, doi: 10.1016/j.xops.2021.100079.
[4]  I. Goodfellow *et al.*, "Generative adversarial networks," *Communications of the ACM,* vol. 63, no. 11, pp. 139-144, 2020.
[5]  P. Andreini *et al.*, "A two-stage gan for high-resolution retinal image generation and segmentation," *Electronics,* vol. 11, no. 1, p. 60, 2022.

[6] P. Costa *et al.*, "End-to-End Adversarial Retinal Image Synthesis," *IEEE Transactions on Medical Imaging,* vol. 37, no. 3, pp. 781-791, 2018, doi: 10.1109/TMI.2017.2759102.

[7] Z. Dong, G. Liu, G. Ni, J. Jerwick, L. Duan, and C. Zhou, "Optical coherence tomography image denoising using a generative adversarial network with speckle modulation," *J Biophotonics,* vol. 13, no. 4, p. e201960135, Apr 2020, doi: 10.1002/jbio.201960135.

[8] T. K. Yoo, J. Y. Choi, and H. K. Kim, "CycleGAN-based deep learning technique for artifact reduction in fundus photography," *Graefe's Archive for Clinical and Experimental Ophthalmology,* vol. 258, no. 8, pp. 1631-1637, 2020.

[9] P. Burlina, N. Joshi, W. Paul, K. D. Pacheco, and N. M. Bressler, "Addressing artificial intelligence bias in retinal diagnostics," *Translational Vision Science & Technology,* vol. 10, no. 2, pp. 13-13, 2021.

[10] A. S. Coyner, J. P. Campbell, J. Kalpathy-Cramer, P. Singh, K. Sonmez, and M. F. Chiang, "Retinal fundus image generation in retinopathy of prematurity using autoregressive generative models," *Investigative Ophthalmology & Visual Science,* vol. 61, no. 7, pp. 2166-2166, 2020.

[11] A. Ha *et al.*, "Deep-learning-based enhanced optic-disc photography," *PloS one,* vol. 15, no. 10, p. e0239913, 2020.

[12] A. Tavakkoli, S. A. Kamran, K. F. Hossain, and S. L. Zuckerbrod, "A novel deep learning conditional generative adversarial network for producing angiography images from retinal fundus photographs," *Scientific Reports,* vol. 10, no. 1, pp. 1-15, 2020.

[13] C. Zheng *et al.*, "Assessment of generative adversarial networks model for synthetic optical coherence tomography images of retinal disorders," *Translational Vision Science & Technology,* vol. 9, no. 2, pp. 29-29, 2020.

[14] S. G. Odaibo, "Generative adversarial networks synthesize realistic OCT images of the retina," *arXiv preprint arXiv:1902.06676,* 2019.

[15] H. Lee, S. Kim, M. A. Kim, H. Chung, and H. C. Kim, "Post-treatment prediction of optical coherence tomography using a conditional generative adversarial network in age-related macular degeneration," *Retina,* vol. 41, no. 3, pp. 572-580, 2021.

[16] Y. Liu *et al.*, "Prediction of OCT images of short-term response to anti-VEGF treatment for neovascular age-related macular degeneration using generative adversarial network," *British Journal of Ophthalmology,* vol. 104, no. 12, pp. 1735-1740, 2020.

[17] H. Cheong *et al.*, "DeshadowGAN: a deep learning approach to remove shadows from optical coherence tomography images," *Translational Vision Science & Technology,* vol. 9, no. 2, pp. 23-23, 2020.

[18] C. Zheng *et al.*, "Assessment of Generative Adversarial Networks for Synthetic Anterior Segment Optical Coherence Tomography Images in Closed-Angle Detection," *Transl Vis Sci Technol,* vol. 10, no. 4, p. 34, Apr 1 2021, doi: 10.1167/tvst.10.4.34.

[19] Z. Zhang *et al.*, "Prediction of corneal astigmatism based on corneal tomography after femtosecond laser arcuate keratotomy using a pix2pix conditional generative adversarial network," *Front Public Health,* vol. 10, p. 1012929, 2022, doi: 10.3389/fpubh.2022.1012929.

[20] R. Gal, D. C. Hochberg, A. Bermano, and D. Cohen-Or, "SWAGAN: A style-based wavelet-driven generative model," *ACM Transactions on Graphics (TOG),* vol. 40, no. 4, pp. 1-11, 2021.

[21] T. Karras, S. Laine, M. Aittala, J. Hellsten, J. Lehtinen, and T. Aila, "Analyzing and improving the image quality of stylegan," in *Proceedings of the IEEE/CVF conference on computer vision and pattern recognition*, 2020, pp. 8110-8119.

[22] M. Lucic, K. Kurach, M. Michalski, S. Gelly, and O. Bousquet, "Are gans created equal? a large-scale study," *Advances in neural information processing systems,* vol. 31, 2018.

[23] N. Chen, A. Klushyn, R. Kurle, X. Jiang, J. Bayer, and P. Smagt, "Metrics for deep generative models," in *International Conference on Artificial Intelligence and Statistics*, 2018: PMLR, pp. 1540-1550.

[24] Y. Shen and B. Zhou, "Closed-form factorization of latent semantics in gans," in *Proceedings of the IEEE/CVF Conference on Computer Vision and Pattern Recognition*, 2021, pp. 1532-1540.

[25] X. Wang *et al.*, "Esrgan: Enhanced super-resolution generative adversarial networks," in *Proceedings of the European conference on computer vision (ECCV) workshops*, 2018, pp. 0-0.

[26] C. Tian, J. Yang, P. Li, S. Zhang, and S. Mi, "Retinal fundus image superresolution generated by optical coherence tomography based on a realistic mixed attention GAN," *Medical Physics,* vol. 49, no. 5, pp. 3185-3198, 2022, doi: https://doi.org/10.1002/mp.15580.

[27] K. Yamashita and K. Markov, "Medical image enhancement using super resolution methods," in *Computational Science–ICCS 2020: 20th International Conference, Amsterdam, The Netherlands, June 3–5, 2020, Proceedings, Part V 20*, 2020: Springer, pp. 496-508.

[28] E. Lyapustin, A. Kirillova, V. Meshchaninov, E. Zimin, N. Karetin, and D. Vatolin, "Towards true detail restoration for super-resolution: A benchmark and a quality metric," *arXiv preprint arXiv:2203.08923,* 2022.

[29] A. J. Sreejith Kumar *et al.*, "Evaluation of Generative Adversarial Networks for High-Resolution Synthetic Image Generation of Circumpapillary Optical Coherence Tomography Images for Glaucoma," *JAMA Ophthalmology,* vol. 140, no. 10, pp. 974-981, 2022, doi: 10.1001/jamaophthalmol.2022.3375.

[30] A. Bakhtiarnia, Q. Zhang, and A. Iosifidis, "Efficient High-Resolution Deep Learning: A Survey," *arXiv preprint arXiv:2207.13050,* 2022.

[31] C. Bowles *et al.*, "Gan augmentation: Augmenting training data using generative adversarial networks," *arXiv preprint arXiv:1810.10863,* 2018.

[32] B. Qiu *et al.*, "Noise reduction in optical coherence tomography images using a deep neural network with perceptually-sensitive loss function," *Biomedical Optics Express,* vol. 11, no. 2, pp. 817-830, 2020.